\documentclass[conference,twocolumn,10pt]{IEEEtran} 

\usepackage{epsfig,cite}
\usepackage{graphicx}
\usepackage{bmpsize}
\usepackage{color}
\usepackage{amssymb,amsmath,mathrsfs}
\definecolor{blue}{rgb}{0,0,1}
\definecolor{darkgreen}{rgb}{0,.5,0}
\definecolor{darkred}{rgb}{.5,0,0}
\newcommand{\red}[1]{\textcolor{black}{#1}}

\IEEEoverridecommandlockouts

\newcommand{\given}{\:|\:}

\def\given{\:|\:}
\def\L{\mathsf{L}}
\def\H{\mathsf{H}}

\usepackage{color}
\renewcommand{\red}[1]{{\color{black} #1}}
\newcommand{\pt}[1]{{\color{black} #1}}
\newcommand{\awe}[1]{{\color{black} #1}}

\def\Pr{{\mathrm{Pr}}}

\def\Pr{{\mathrm{Pr}}}

\begin{document}

\title{Linear Noise Approximation of Intensity-Driven Signal Transduction Channels}

\author{
	\IEEEauthorblockN{Gregory R.~Hessler$^1$, Andrew W.~Eckford$^2$, and Peter J.~Thomas$^3$}
	%
	%
	\IEEEauthorblockA{$^{1,3}$ Dept. of Math., Appl. Math., and Stats., Case Western Reserve University, Cleveland, OH, USA\\
	$^2$Dept. of EECS, York University, Toronto, ON, Canada\\
    Emails:  $^1$gregory.hessler@case.edu, $^2$aeckford@yorku.ca, $^3$pjthomas@case.edu}
	\thanks{This work was supported by a grant from the Natural Sciences and Engineering Research Council to AWE and National Science Foundation grants DMS-1413770 and DEB-1654989 to PJT.}}

\maketitle

\begin{abstract}
Biochemical signal transduction, a form of molecular communication, can be modeled using graphical Markov channels with input-modulated transition rates.  Such channel models are strongly non-Gaussian.   In this paper we use a linear noise approximation to construct a novel class of Gaussian additive white noise channels that capture essential features of fully- and partially-observed intensity-driven signal transduction.  When channel state transitions that are sensitive to the input signal are directly observable, high-frequency information is transduced more efficiently than low-frequency information, and the mutual information rate per bandwidth (spectral efficiency) is significantly greater than when sensitive transitions and observable transitions are disjoint.  When both observable and hidden transitions are input-sensitive, we observe a superadditive increase in spectral efficiency.
\end{abstract}

\section{Introduction}

Communication within biological systems often involves chemical processes in which a receptor (protein, DNA or a living cell) transitions among several metastable states, with transitions rates modulated by the intensity of an input signal.  Examples of such input signals include: the concentration of a ligand, transcription factor, or metabolic substrate; intensity of a light source; and potential difference across a cell membrane.  

A simple example of such a channel is the BIND channel \cite{ThomasEckford2016IEEE_TransIT} in which a receptor protein has two states: bound to the ligand, or unbound; the receptor is sensitive to the signal (ligand concentration) in the unbound state, and insensitive in the bound state.  
More elaborate channel schemes abound in biology: examples include the Channelrhodopsin-2 receptor (ChR2) \cite{nag03} and the acetylcholine receptor (ACh) \pt{\cite{sakmann1980single,Colquhoun+Hawkes:1983chapter}}, with three and five states, respectively. These systems are not restricted to receptors in signal transduction: similar models have been proposed for transcription of RNA in the presence of a promoter \cite{rieckh2014}. 

Receptor \pt{proteins} may be modelled as Markov processes on a finite graph, with vertices representing the state of the receptor, and a weighted directed edge $a\rightarrow b$, with weight representing the transition rate from state $a$ to state $b$. In general, not all of these edges have rates that are sensitive to the input signal, and not all the state transitions are observable by the organism. 
We will describe these systems more formally in the next section, but from this description the system output can also be modelled as a {\em hidden Markov chain}.  
\pt{
\begin{equation}\label{diag:xyz}
\begin{array}{ccccccccc}
&X_1&&X_2&&X_3&\cdots\\
&\downarrow&&\downarrow&&\downarrow&&  \\
(Y_0)&\longrightarrow&Y_1&\longrightarrow&Y_2&\longrightarrow&Y_3&\longrightarrow&\cdots\\
&&\downarrow&&\downarrow&&\downarrow   \\
&&Z_1&&Z_2&&Z_3&\cdots
\end{array}\end{equation}
Diagram~\eqref{diag:xyz} illustrates the structure of such a partially observed channel model, in a discrete time formulation: A sequence of input signals ($X_1,X_2,\ldots)$ modulates the transition rates  governing the evolution of a channel state vector ($Y_1,Y_2,\ldots$), a subset of which form the observable channel output ($Z_1,Z_2,\ldots$).  The {\em fully observed} channel refers to $X\to Y$; the {\em partially observed} channel to $X\to Z$.
}

There are many ways to evaluate the performance and capabilities of a signalling system, including ideal observer and signal detection analysis \cite{swets2014signal} and fidelity of stimulus reconstruction \cite{berger1998}.
In this paper we consider the framework of information theory, particularly the mutual information \pt{(MI)} between the input signal and the state of the receptor (either all states, \pt{$Y$,} or the subset of observable states, \pt{$Z$}). In our previous work, we calculated mutual information for individual receptors including ChR2 and ACh \cite{EckfordThomas2019IEEE_TMBMC}, and the BIND receptor for multiple receptors \cite{ThomasEckford2016ISIT}; elsewhere, mutual information has been used as a figure of merit in biological systems such as the sensitivity of gene expression \cite{tkavcik2008information,tkavcik2011information}. 

\begin{figure*}[h!]
    \centering
    \includegraphics[width=\textwidth]{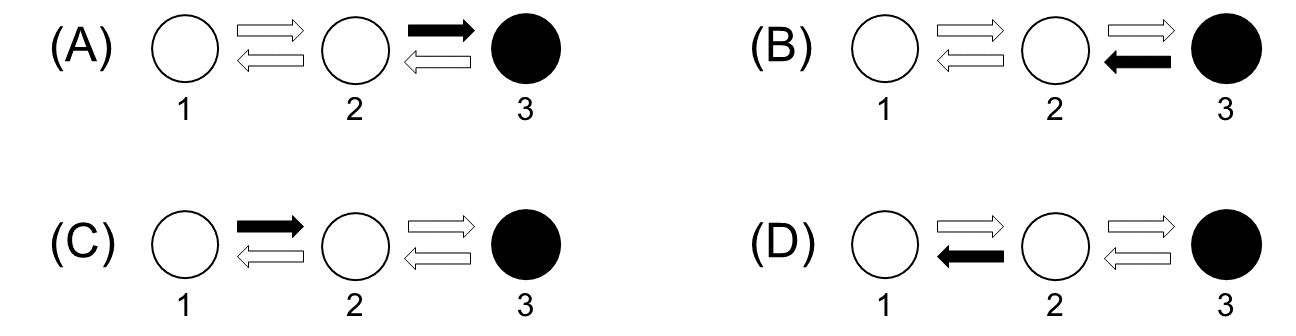}
    \caption{\pt{Structure of a partially-observed intensity-modulated Markov channel with three states.  The channel state is $Y=[Y_1,Y_2,Y_3]$; the channel output is $Z=[0,0,1]\cdot Y=Y_3$. Thus transitions along edge $2\to3$ and $3\to2$ are observable, while transitions $1\to 2$ and $2\to 1$ are hidden.  In each version of the model, a single edge is sensitive to the input put $X$ (filled arrows) and the remaining edges are insensitive (open arrows). In A and B the sensitive edge is observable; In C and D the sensitive edge is hidden.  Qualitatively, we expect the mutual information $I(X;Z)$ to be larger for A and B than for C and D.}
}
    \label{fig:3-state-model-four-versions}
\end{figure*}

\awe{Since the partially observed system consists of the features of signal transduction that are used by the organism, a complete analysis of the partially observed system is essential. While the fully observed system MI can be obtained in closed form for IID inputs \cite{EckfordThomas2019IEEE_TMBMC}, \red{for the partially observed system the MI is generally intractable. Explicit} expressions for the entropy rates of hidden Markov chains are often difficult to obtain, \red{and new, approximate methods are required.}

Our previous investigations revealed a qualitative relationship between the structure of \emph{sensitivity} (which state transitions are modulated by the input) and \emph{observability} (which components of $Y$ are included in $Z$) and the MI of the fully and partially observed systems.}
In some cases the fully observed system has the same, or nearly the same MI as the partially observed system, so the latter can be studied using the closed form expression for the former.  In other cases, such as the important neurotransmitter ACh, the fully observed and partially observed MI differ by an order of magnitude.

\awe{The key contribution of this paper is to} \pt{develop a linear noise approximation \cite{elf2003fast,wallace2012linear} applicable to  large populations of senders and receivers, that gives a closed-form solution for the spectral efficiency (MI per bandwidth) of both the fully- and partially-observed signal transduction channel models.}
\pt{For our linear channel approximation, we} show that high frequency information is transduced more efficiently when observable transitions are also sensitive; we also show that the spectral efficiency of receptors is dependent on whether sensitive transitions are observable. These results have biological significance, for example in predicting how an organism will observe high-frequency phenomena.

\section{System model}

\subsection{Physical model}

Consider a biochemical receptor that is sensitive to a given input process $X(t)$.  \pt{We first describe a discrete-time channel model, $t\in\Delta t\cdot \mathbb{N}$, from which we will derive a continuous time model expressed as a stochastic differential equation \eqref{eq:SDE_with_input} with $t\in\mathbb{R}$.}
As noted in the introduction, different types of receptors are sensitive to a variety of input stimuli.

\awe{{\bfseries For concreteness, here we will consider a simple three-state chain with one observable state \pt{(cf.~Fig.~\ref{fig:3-state-model-four-versions})}.}  (However, our framework can be extended to general signal transduction systems, such as those considered in \cite{EckfordThomas2019IEEE_TMBMC}.) We see that:}
\begin{itemize}
	\item {\em Only some of the state transitions are sensitive to the input.} \awe{In Fig.~\ref{fig:3-state-model-four-versions}, only the filled arrows are sensitive to the concentration of the ligand: their transition rate is directly proportional to concentration. The unfilled arrows always occur at the same rate, independent of concentration.}
	
	\item {\em Only some of the state \awe{transitions} are observable by the organism.} \awe{In Fig.~\ref{fig:3-state-model-four-versions}, the unfilled bubbles (representing states 1 and 2) are indistinguishable to the organism. Thus, only transitions between states 2 and 3 are observable, while transitions between 1 and 2 occur internally to the receptor, hidden from the organism. Physically, this may occur if the receptor has an ion channel, which is closed in states 1 and 2, and open in state 3.}

\end{itemize}

We consider both ``fully observed" systems (the channel with signal $X(t)$ as input and channel state $Y(t)$ as output) and  ``partially observed" systems (the channel with $X(t)$ as input and $Z(t)=C^\intercal Y(t)$ as output, where $C$ is a rectangular matrix giving the observable output).
Intuitively, we expect most of the information transduced about the signal to come from the edges that are both observable and sensitive.  But even in networks for which all sensitive edges are hidden, there can still be a positive mutual information and capacity.  



We consider an input signal generated by an ensemble of $M\gg 1$ independent binary sources that together produce a concentration or intensity $c(t)=c_{\L}+\Delta c\, V(t)/M$, where $\Delta c=(c_\H-c_\L)$, $V(t)\sim\text{Binom}(M,p)$, and $0\le c_{\L}<c_{\H}$ are the low and high concentration signals that would obtain were all $M$ sources simultaneously inactive, or active, respectively. 
As a simplification, we assume the input process is uncorrelated across discrete timesteps of size $\Delta t$ (see Discussion).
\pt{We let $\epsilon=1/\sqrt{M}$ represent the relative size of fluctuations in the input concentration signal. \red{By the central limit theorem,} for sufficiently large $M$}, we may approximate \pt{the input signal} $c(t)$ with a \pt{discrete time} Gaussian white noise input, and write the input process $X(t)$ as
\begin{equation}
    X(t)=c(t)-\overline{c} = \epsilon \, \Delta c\, \sqrt{p(1-p)}\xi(t)
\end{equation}
where $\overline{c}=c_\L+p\Delta c$ is the mean concentration, the variance of the input is $\left(\Delta c\right)^2 p(1-p)/M$, 
$\xi$ is Gaussian white noise with $\langle\xi(t)\rangle=0$, \pt{and} $\langle\xi(t)\xi(t')\rangle=\delta(t-t')$.
As a rule of thumb we restrict attention to parameter values for which the Gaussian approximation is appropriate, namely $10<Mp<(M-10)$. For the systems we consider, $M\gtrsim 1000$, so $.01<p<.99$ is a suitable parameter range.

Next, consider a population of $N\gg 1$ independent receptors, each \pt{individually} described by a channel state \pt{given by one of the standard unit vectors $Y(t)\in\mathcal{S}\equiv \{\mathbf{e}_1,\mathbf{e}_2,\mathbf{e}_3\}$.}
We have shown previously that for a population of independent binary-state BIND receptors \pt{receiving common IID input}, the mutual information and capacity is exactly $N$ times the single-receptor mutual information and capacity, respectively \cite{ThomasEckford2016ISIT}.  
An analogous result holds for arbitrary finite-state receptors with IID input (\red{the proof is straightforward but lengthy; we omit it for space}).
We assume that each receptor acts, independently of the others, as a conditional Markov process on $\mathcal{S}$, with transition rates depending on the input concentration $c=\overline{c}+X(t)$.  
Assuming standard first-order mass-action kinetics \cite{BriggsHaldane1925BiochemJ}, the transition probability from state $i$ to state $j$ is affine-linear.  Thus for each receptor we have
\begin{align}
\Pr[Y(t&+\Delta t)=j\given Y(t)=i,X(t)=x]\\ \nonumber 
=&\left(\alpha_{ij}+\beta_{ij}x\right)\Delta t+o(\Delta t),\text{ as }\Delta t\to 0.
\end{align}
%
%
Recall that only some of the state transitions are sensitive to the input concentration. If $\beta_{ij}=0$, we \red{define} the $i\to j$ transition \red{to be} \emph{insensitive}; otherwise it is \emph{sensitive}.  
For the sensitive transitions, the mean input concentration affects the mean transition rate.  Thus we have $\alpha_{ij}(p)=k_{ij}+c_\L\beta_{ij}+p\beta_{ij}\Delta c,$ where $k_{ij}$ is the intrinsic $i\to j$ transition rate.  
(Introducing a nonlinear dependence of the $\alpha_{ij}$ on $p$ \pt{would} not qualitatively change our results.)
We will require that the channel state process is irreducible for all  $0<p<1$.

\pt{Just as we assume the population of independent transmitters is large enough to justify a Gaussian approximation for the input concentration signal, we also assume a large population of $N\gg1$ independent receptors exposed to the same input signal.
For large $N$, the fluctuations in number of transitions per time among states, \red{relative to the mean number of transitions per unit time}, scale as $1/\sqrt{N}$ \cite{SchmidtGalanThomas2018stochastic}.  In general, we may consider a variety of scaling relations between $M$, the number of sources, and $N$, the number of receptors.  For this paper we assume $M=N$ for simplicity.  
Our results would not change significantly with $N\propto M$ in some other fixed proportion.  
Thus we consider $N=M$} independent receptors all exposed to an identical input concentration $c(t)=\overline{c}+X(t)$.  \red{Kurtz's theorem guarantees that (over any given finite time horizon) the discrete process converges as $N\to\infty$ to a Gaussian process \cite{Kurtz1971JApplProb}. Thus for sufficiently large $N$, we may} approximate the receptor state \textit{via} a chemical Langevin equation \cite{Gardiner2004,SchmidtThomas2014JMN}.
We write the three-state \pt{ fractional population vector  as} $Y(t)=(Y_1,Y_2,Y_3)^\intercal$ with $Y_1+Y_2+Y_3\equiv 1$ \pt{and $Y_i\ge 0$}.
\pt{To derive a tractable channel model based on the linear noise approximation, we begin with the exact (discrete) representation with $N$ channels, in which case}
 $Y_i\in\frac1N\{0,1,2,\ldots,N\}$.  For \pt{$N\gg 1$}, 
we \pt{may approximate} the population state vector $Y(t)$ as a Gaussian random vector process in continuous time, \pt{based on several observations:}
%

\pt{
\begin{enumerate}
\item In the absence of input fluctuations ($c(t)=\overline{c}$) the mean channel state obeys a linear evolution equation with stable eigenvalues (except the Perron-Frobenius eigenvalue corresponding to the steady state channel distribution).  
\item The input fluctuations are stationary, so $Y$  has a stationary distribution,
given by
$\overline{Y}_1=\alpha_{32}\alpha_{21}/\mathcal{Z},\quad
\overline{Y}_2=\alpha_{32}\alpha_{12}/\mathcal{Z},\quad
\overline{Y}_3=\alpha_{12}\alpha_{23}/\mathcal{Z},$, with $
\mathcal{Z}=\alpha_{32}\alpha_{21}+\alpha_{32}\alpha_{12}+\alpha_{12}\alpha_{23}.$
Generally both $\overline{Y}$ and the normalization  $\mathcal{Z}$ depend on $p$.
\item For $M\gg 1$, the fluctuating part of the input concentration signal $\overline{c}+X(t)$ will be small (variance $O(1/M)$).
\item The variance of the channel state, $E[(Y_i-\overline{Y}_i)^2]$, will be small when both $M$ and $N$ are large.  
\end{enumerate}
}
To obtain a linear noise approximation\cite{wallace2012linear}, 
\pt{we expand the joint $(X,Y)$ process around the mean, and}
neglect terms  
smaller than size $O(\epsilon)$.  
We subtract the mean with a linear change of coordinates $Y\to Y-\overline{Y}$.
Thus, $Y$ represents the deviations of the channel population vector from its mean value (therefore  $E[Y]=0$).  We thereby obtain a first-order approximation with additive Gaussian noise
\pt{represented by the increments of several independent Wiener processes ($\Delta W_{ij}$ for the intrinsic fluctuations in the $i\to j$ transition):}
\begin{align}
Y_1(t+\Delta t)&-Y_1(t)=
\label{eq:discrete_time_Gaussian_linear1}
\\ \nonumber
&-\left(\left( \overline{Y}_1\beta_{12}X+Y_1\alpha_{12}\right)\Delta t
+\epsilon\sqrt{\overline{Y}_1\alpha_{12} \Delta t}\,W_{12}\right)\\
\nonumber
&+\left(\left( \overline{Y}_2\beta_{21}X+Y_2\alpha_{21}\right)\Delta t
+\epsilon\sqrt{\overline{Y}_2\alpha_{21} \Delta t}\,W_{21}\right)\\
\label{eq:discrete_time_Gaussian_linear3}
Y_3(t+\Delta t)&-Y_3(t)=\\ \nonumber 
&-\left(\left( \overline{Y}_3\beta_{32}X+Y_3\alpha_{32}\right)\Delta t
+\epsilon\sqrt{\overline{Y}_3\alpha_{32} \Delta t}\,W_{32}\right)\\
\nonumber
&+\left(\left( \overline{Y}_2\beta_{23}X+Y_2\alpha_{23}\right)\Delta t
+\epsilon\sqrt{\overline{Y}_2\alpha_{23} \Delta t}\,W_{23}\right)\\
Y_2(t+\Delta t)&=1-(Y_1(t+\Delta t)+Y_3(t+\Delta t)).
\label{eq:discrete_time_Gaussian_linear2}
\end{align}
\normalsize
Writing the system more compactly, equations \eqref{eq:discrete_time_Gaussian_linear1}-\eqref{eq:discrete_time_Gaussian_linear2} have the form
\small 
\begin{equation}
\mathbf{Y}(t+\Delta t)=(I+A\Delta t) \mathbf{Y}(t)+\mathbf{b}\,\Delta t\,X(t)+\epsilon\,\sqrt{\Delta t}\,\Gamma\,\mathbf{W}(t)
\end{equation}
\normalsize
where the $3\times 3$ matrix $A$, the $3\times 1$ vector $\mathbf{b}$, and the $3\times 4$ matrix $\Gamma$ all depend on the parameter $p$, and $\mathbf{W}$ is a $4\times 1$ vector of standard Gaussian random variables, with independent components and uncorrelated in time.
Thus we arrive at a linear channel model, with structure arising from the channel state transition graph.  The input signal is $X(t)$, a mean-zero Gaussian discrete time process with variance $\epsilon^2 p(1-p)(\Delta c)^2$, the channel noise is given by the Wiener processes comprising $\mathbf{W}$, and the channel memory arises from the mean transition rate matrix $A$.   

In order to obtain analytical results we will consider a continuous time limit in which the channel model takes the form of a multidimensional Ornstein-Uhlenbeck process \cite{Gardiner2004}
\begin{align}\label{eq:SDE_with_input}
d\mathbf{Y}&=A\mathbf{Y}\,dt+\epsilon \mathbf{b}\,\sqrt{p(1-p)}\,\Delta c \,dW_x+\epsilon\mathcal{B}\,d\mathbf{W}.
\end{align}
Here $\mathbf{Y}$ is a vector with components summing to zero, representing the fluctuations of the channel state around the mean occupancy (with the mean occupancy normalized to unity); $A$ is a $3\times 3$ matrix (depending on $p$) with a single null eigenvector corresponding to the stationary distribution, and remaining eigenvalues real and negative.  
The independent Wiener process increments $dW_x(t)$ represents the input.  
The input fluctuations are ``transduced" according to the vector $\mathbf{b}=\sum_{k\in\mathcal{E}^*}\beta_{ij(k)}\zeta_k$, where \pt{$\zeta_k=-\mathbf{e}_{i(k)}+\mathbf{e}_{j(k)}$} is the stoichiometry vector for directed edge $k:i(k)\to j(k)$ and $\mathcal{E}^*$ is the collection of sensitive edges in the graph.  Note that the components of $\mathbf{b}$ sum to zero.  
The channel noise arises from the vector $d\mathbf{W}$ of independent Wiener process increments; note $d\mathbf{W}$ is also independent of $dW_x$.
For the three-state model with four edges, $\mathcal{B}$ is a $3\times 4$ matrix
\begin{equation}
    \mathcal{B}=\left(\begin{array}{c|c|c|c}
    \sqrt{J_1}\zeta_1 & \sqrt{J_2}\zeta_2 & \sqrt{J_3}\zeta_3 & \sqrt{J_4}\zeta_4
    \end{array}\right).
\end{equation}
Here $J_i=\overline{Y}_i\alpha_{ij(k)}$ is the mean flux across the $k$th directed edge, with $ij(k)$ denoting the pair $i\to j$ corresponding to edge $k$. 
A similar construction results for arbitrary signal-transduction systems represented as a directed graph $(\mathcal{V},\mathcal{E})$ with $|\mathcal{V}|$ vertices and $|\mathcal{E}|$ edges; in general $\mathcal{B}$ will be $|\mathcal{V}|\times |\mathcal{E}|.$

The mutual information rate between the Gaussian input and multidimensional Gaussian output processes $X(t)$ and $Y(t)$ in \eqref{eq:SDE_with_input} is given by the integral of the spectral efficiency \cite{kolmogorov1956shannon,pinsker1964information,ihara1993information}
\begin{align}\label{eq:SE-fully-observed}
&\text{SE}^\text{FULL}(\omega)=\frac12\log\left(\mathcal{R}(\omega)\right)\\    \nonumber 
&\mathcal{R}=\frac{
    	\det^*\Big((A + i \omega)^{-1} (\mathcal{B} \mathcal{B}^T + p (1-p)\mathbf{b}\mathbf{b}^T) (A^T - i \omega)^{-1}\Big)
   }
   {
   		\det^*\Big((A + i \omega)^{-1} \mathcal{B} \mathcal{B}^T (A^T - i \omega)^{-1}	\Big)
   }.
\end{align}
In this formula $\det^*$ denotes the pseudodeterminant, the product of the nonzero eigenvalues of a square matrix.  
For $|\omega|>0$ the matrix $(A\pm i\omega I)$ is full rank; the matrix $\mathcal{B}\mathcal{B}^\intercal$ has rank $|\mathcal{V}|-1$.  
The rank-one matrix $\mathbf{b}\mathbf{b}^\intercal$ is the outer product of a vector $\mathbf{b}$ that is a linear combination of the columns of $\mathcal{B},$ each of which is proportional to one of the stoichiometry vectors $\zeta_k$.  
It follows that the matrix $\mathcal{B}\mathcal{B}^\intercal + p(1-p)\mathbf{b}\mathbf{b}^\intercal$ has the same rank as $\mathcal{B}\mathcal{B}^\intercal$.  The matrices  in \eqref{eq:SE-fully-observed} are Hermitian and have the same number of nonzero (and real) eigenvalues, \pt{thus} the ratio of pseudodeterminants is well defined. 
Note that the factor of $\epsilon$ multiplying $\mathbf{b}$ and $\mathcal{B}$ in \eqref{eq:SDE_with_input} cancels in the ratio $\mathcal{R}$. Therefore, \pt{provided $\epsilon\ll 1$ is small enough (that is, $M,N\gg 1$ are large enough) to ensure validity of the underlying Gaussian approximations, the following results hold independent of the specific value of $\epsilon.$}
\begin{figure*}[ht!]
\centering
\includegraphics[width=0.9\textwidth]{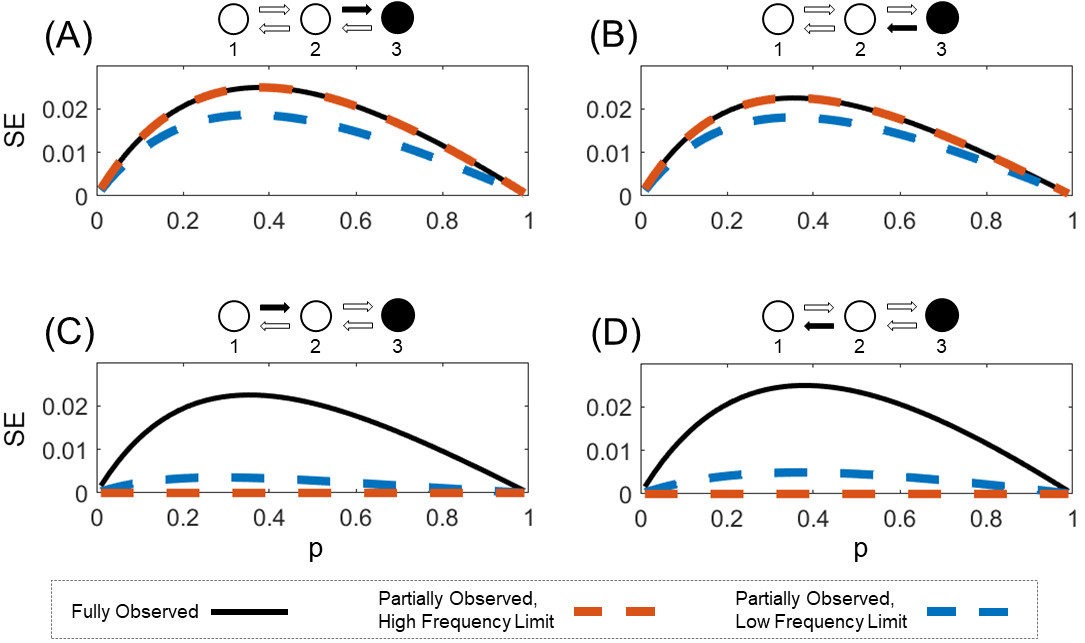}
\caption{Spectral efficiency (SE, (bits/sec)/Hz) as a function of input parameter $p$, when the system is fully observable (solid black line) or when only \awe{the transition to} state 3 is observable (dashed lines).   Under partial observability, SE is frequency-dependent.  Red (resp.~blue) dashing shows the high (resp.~low) frequency SE limits.  In each panel a single transition is sensitive (filled arrow) to the input, and three transitions are insensitive (open arrows). \textbf{A:} Observable transition $2\to 3$ is sensitive.  \textbf{B:} Observable transition $3\to 2$ is sensitive. \textbf{C:} Hidden transition $1\to 2$ is sensitive. \textbf{D:} Hidden transition $2\to 1$ is sensitive.}
\label{fig:single_edge_sensitive}
\end{figure*}

We wish to compare the spectral efficiency for the fully-observed system \eqref{eq:SE-fully-observed} with that of the partially-observed system.
Let the observation vector $C=(0,0,1)^\intercal$ (meaning transitions $2\leftrightharpoons 3$ are directly observable, while transitions $1\leftrightharpoons 2$ are not).  Then the spectral efficiency appearing in the expression for the mutual information rate between the input $X(t)$ and the observed channel state $Z(t)=C^\intercal Y(t)$ satisfies
\begin{align}
    &\text{SE}^\text{PART}(\omega)=\frac12\log\left(\mathcal{Q}(\omega)\right)\\ \nonumber
    &\mathcal{Q}=
    \frac{
    	C^\intercal (A + i \omega)^{-1} (\mathcal{B} \mathcal{B}^T +  p (1-p)\mathbf{b}\mathbf{b}^T) (A^T - i \omega)^{-1}C
   }
   {
   		C^\intercal (A + i \omega)^{-1} \mathcal{B} \mathcal{B}^T (A^T - i \omega)^{-1}	C
   }.
\end{align}
In general, the integrals $\int_{|\omega|<\omega_\text{max}}\log\mathcal{R}(\omega)\,d\omega$ and $\int_{|\omega|<\omega_\text{max}}\log\mathcal{Q}(\omega)\,d\omega$ diverge as $\omega_\text{max}$ grows without bound.  Rather than impose an arbitrary cutoff frequency, we study the integrands directly in their high- and low-frequency limits, 
\begin{align}
    \text{SE}_0\equiv\lim_{\omega\to 0}\text{SE}(\omega);\quad \quad \text{SE}_\infty\equiv \lim_{|\omega|\to\infty}\text{SE}(\omega).
\end{align}
While $\text{SE}^\text{PART}(\omega)$ varies over a finite range as $\omega$ varies $0\le |\omega|<\infty$, we find that $\text{SE}^\text{FULL}$ is independent of $\omega$ and can be simplified to
\begin{equation}\label{eq:SE-fully-observed-frequency-independent}
\text{SE}^\text{FULL}(\omega)=\frac12\log\left(\frac{
    	\det^*\Big( \mathcal{B} \mathcal{B}^T + p (1-p)\mathbf{b}\mathbf{b}^T \Big)
   }
   {
   		\det^*\Big(\mathcal{B} \mathcal{B}^T 	\Big)
   }\right)\\    
.
\end{equation}

\section{Results}

\begin{figure*}[htbp]
\centering
\includegraphics[width=0.9\textwidth]{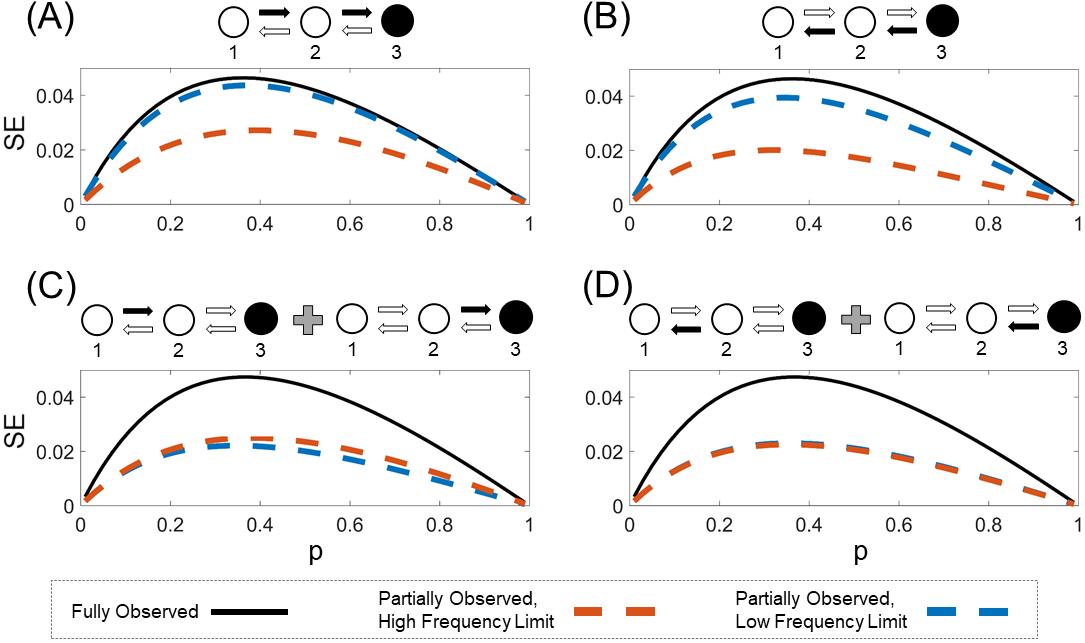}
\caption{Spectral efficiency (SE, (bits/sec)/Hz) as a function of input parameter $p$, when two transitions are sensitive and two insensitive.  Conventions as in Fig.~\ref{fig:single_edge_sensitive} (black line: fully observed; red dashed line: partially observed, high-frequency limit; blue dashed line: partially observed, low-frequency limit).
\textbf{A:} Hidden transition $1\to 2$ and observable transition $2\to 3$ are sensitive.  \textbf{B:} Hidden transition $2\to 1$ and observable transition $3\to 2$ are sensitive. \textbf{C:} Sum of SE from Figs.~2A ($2\to 3$ sensitive) and 2C ($1\to 2$ sensitive).  \textbf{D:} Sum of SE from Figs.~2B ($3\to 2$ sensitive) and 2D ($2\to 1$ sensitive).}
\label{fig:two_edges_sensitive}
\end{figure*}

Figs.~\ref{fig:single_edge_sensitive}-\ref{fig:two_edges_sensitive} shows the spectral efficiency (SE) of the fully observed system (solid black lines), together with the spectral efficiency of the partially observed system in the high-frequency (dashed red lines) and low-frequency (dashed blue lines) limits, as the  high-concentration input probability $p$ ranges from 0.01 to 0.99.  
For all systems we use standard parameters $\alpha_{ij}=1, \beta_{ij}=0$ for insensitive edges and $\alpha_{ij}=1,\beta_{ij}=1$ for sensitive edges. Inset diagrams indicate which edges are sensitive (solid black arrows), resp.~insensitive (open arrows), \pt{following the same scheme as Fig.~\ref{fig:3-state-model-four-versions}}.

In Fig.~\ref{fig:single_edge_sensitive}, a single transition is sensitive to the input.  When this transition is observable, the high-frequency SE is indistinguishable from the fully-observed SE (Fig.~2A, 2B, black and red dashed lines), while the low-frequency SE is smaller (blue dashed lines).  In contrast, when only the observable transitions are insensitive, the low-frequency SE is reduced, but the high-frequency SE is effectively zero (Fig.~2C, 2D). Thus, heuristically,  only the slow time scale carries information about the hidden transition.  

In each case the maximal value of SE occurs at an interior value of $p$, ranging roughly from 0.3 to 0.4.  The maximal SE for the fully observed system is 0.025 for cases A and D (which are identical, by symmetry, for the parameters we use) and is 0.022 for cases B and C.  


Fig.~\ref{fig:two_edges_sensitive} contrasts a system with two sensitive transitions directed towards the third state (the transition to which is observable) (Fig.~3A), and a system with two  sensitive transitions directed away from the third state (Fig.~3B).  For the fully observed system, these two cases are identical and hence have identical spectral efficiency curves.  However, for the partially observed system, we find that both the low and high frequency spectral efficiency limits are higher for the system with sensitive transitions moving towards state 3 than the system with sensitive transitions moving away from state 3.  In contrast to the systems with a single sensitive edge, we find that the spectral efficiency at low frequencies exceeds that at high frequencies by a significant margin, when two sensitive edges operate in concert.  

Panel 3C shows the sum of the two SE curves for the systems with a single edge directed towards the third state (sum of curves in 2A and 2C).  Panel 3D shows the sum of the SE curves for the systems with a single edge directed away from the third state (sum of curves in 2B and 2D).  

The spectral efficiency of the fully observed system with edges $1\to 2\to 3$ sensitive is within 5\% of the sum of the corresponding single-edge SE curves (compare Fig.~3A, 3C).  The SE of the fully observed system with sensitive edges $3\to 2\to 1$ is within 12\% of the sum of the $3\to 2$ and $2\to 1$  systems.
The mechanism leading to higher SE for low- versus high-frequency inputs (Figs.~3A-B) remains to be explored.

\section{Discussion}
By considering a large population of independent sources signaling via concentration fluctuations, and a large population of independent \pt{protein} molecules transducing the concentration signal through intensity-mediated state changes, we derived an additive Gaussian channel based on a linear noise approximation. Although models we have studied recently are more realistic for single receptors \cite{nips:ThomasEtAl:2004,ThomasEckford2016IEEE_TransIT,ThomasEckford2016ISIT,EckfordThomas2019IEEE_TMBMC}, the model in this paper is analytically tractable for large populations with hidden states, and shows how signal transduction depends on the structure of observability and sensitivity in a channel.
 For example,
Figs.~2A and 2C suggest that directly observable transitions carry information about the input on both fast and slow timescales, while hidden transitions carry information on slow timescales.  This reflects the stochastic shielding phenomenon \cite{SchmandtGalan2012PRL,SchmidtThomas2014JMN}; cf.~Fig.~4A of \cite{SchmidtGalanThomas2018stochastic}; see also \cite{BernardiLindner2014JNPhys}.

\pt{A shortcoming of the present treatment is the assumption of unlimited input signal power.  
Moreover, rapid signaling via molecular concentrations is limited by the  physics of diffusion, and a more realistic treatment would involve an input with a natural high frequency cutoff. 
The typical persistence time $\tau$ of a signaling molecule with diffusion constant $D$, in a geometry of size $L$, is roughly $\tau\sim L^2/2D.$  For acetylcholine signaling in the neuromuscular junction (NMJ), typical values are $D=4.0\times 10^{-4}\mu\text{m}^2/\mu\text{s}$ and $L\lesssim 1\mu\text{m}$ \cite{tai2003finite}, giving $\tau\lesssim 1$ msec.  
For comparison, nerve cells signaling through the NMJ transmit action potentials with maximal rates in the tens of Hz \cite{bigland1983changes}. 
In \cite{ThomasEckford2016ISIT} we considered the two-state BIND model driven both with IID inputs and with inputs having finite correlation time, and in \cite{eckford2018thermodynamic} we considered  thermodynamic constraints on diffusion-based signaling schemes.  Further analysis along these lines would be a natural next step for the linear channel model explored here.
}

\vspace{-5mm}

\bibliographystyle{ieeetr}
\bibliography{math,PJT,awe_extra,infotheory,signaling,stoch_chem}

\begin{thebibliography}{10}

\bibitem{ThomasEckford2016IEEE_TransIT}
P.~J. Thomas and A.~W. Eckford, ``Capacity of a simple intercellular signal
  transduction channel,'' {\em IEEE Transactions on Information Theory},
  vol.~62, pp.~7358--7382, Dec 2016.

\bibitem{nag03}
G.~Nagel, T.~Szellas, W.~Huhn, S.~Kateriya, N.~Adeishvili, P.~Berthold,
  D.~Ollig, P.~Hegemann, and E.~Bamberg, ``Channelrhodopsin-2, a directly
  light-gated cation-selective membrane channel,'' {\em PNAS}, vol.~100,
  no.~24, pp.~13940--13945, 2003.

\bibitem{sakmann1980single}
B.~Sakmann, J.~Patlak, and E.~Neher, ``Single acetylcholine-activated channels
  show burst-kinetics in presence of desensitizing concentrations of agonist,''
  {\em Nature}, vol.~286, no.~5768, p.~71, 1980.

\bibitem{Colquhoun+Hawkes:1983chapter}
D.~Colquhoun and A.~G. Hawkes, {\em Single-Channel Recording}, ch.~The
  Principles of the Stochastic Interpretation of Ion-Channel Mechanisms.
\newblock Plenum Press, New York, 1983.

\bibitem{rieckh2014}
G.~Rieckh and G.~Tka\v{c}ik, ``Noise and information transmission in promoters
  with multiple internal states,'' {\em Biophys. J.}, vol.~106, pp.~1194--1204,
  2014.

\bibitem{swets2014signal}
J.~A. Swets, {\em Signal detection theory and ROC analysis in psychology and
  diagnostics: Collected papers}.
\newblock Psychology Press, 2014.

\bibitem{berger1998}
T.~Berger and J.~D. Gibson, ``Lossy source coding,'' {\em IEEE Transactions on
  Information Theory}, vol.~44, pp.~2693--2723, 1998.

\bibitem{EckfordThomas2019IEEE_TMBMC}
A.~W. Eckford and P.~J. Thomas, ``The channel capacity of channelrhodopsin and
  other intensity-driven signal transduction receptors,'' {\em IEEE
  Transactions on Molecular, Biological, and Multi-Scale Communications}, 2019.
\newblock In press (arXiv preprint arXiv:1804.04533).

\bibitem{ThomasEckford2016ISIT}
P.~J. Thomas and A.~W. Eckford, ``Shannon capacity of signal transduction for
  multiple independent receptors,'' in {\em 2016 IEEE International Symposium
  on Information Theory (ISIT)}, pp.~1804--1808, July 2016.

\bibitem{tkavcik2008information}
G.~Tka{\v{c}}ik, C.~G. Callan, and W.~Bialek, ``Information flow and
  optimization in transcriptional regulation,'' {\em Proceedings of the
  National Academy of Sciences}, 2008.

\bibitem{tkavcik2011information}
G.~Tka{\v{c}}ik and A.~M. Walczak, ``Information transmission in genetic
  regulatory networks: a review,'' {\em Journal of Physics: Condensed Matter},
  vol.~23, no.~15, p.~153102, 2011.

\bibitem{elf2003fast}
J.~Elf and M.~Ehrenberg, ``Fast evaluation of fluctuations in biochemical
  networks with the linear noise approximation,'' {\em Genome research},
  vol.~13, no.~11, pp.~2475--2484, 2003.

\bibitem{wallace2012linear}
E.~Wallace, D.~Gillespie, K.~Sanft, and L.~Petzold, ``Linear noise
  approximation is valid over limited times for any chemical system that is
  sufficiently large,'' {\em IET systems biology}, vol.~6, no.~4, pp.~102--115,
  2012.

\bibitem{BriggsHaldane1925BiochemJ}
G.~E. Briggs and J.~B.~S. Haldane, ``A note on the kinetics of enzyme action,''
  {\em Biochemical journal}, vol.~19, no.~2, p.~338, 1925.

\bibitem{SchmidtGalanThomas2018stochastic}
D.~R. Schmidt, R.~F. Gal{\'a}n, and P.~J. Thomas, ``Stochastic shielding and
  edge importance for markov chains with timescale separation,'' {\em PLoS
  computational biology}, vol.~14, no.~6, p.~e1006206, 2018.

\bibitem{Kurtz1971JApplProb}
T.~G. Kurtz, ``Limit theorems for sequences of jump markov processes
  approximating ordinary differential processes,'' {\em Journal of Applied
  Probability}, vol.~8, no.~2, pp.~344--356, 1971.

\bibitem{Gardiner2004}
C.~W. Gardiner, {\em Handbook of Stochastic Methods for Physics, Chemistry, and
  the Natural Sciences}.
\newblock Springer Verlag, 2nd~ed., 2004.

\bibitem{SchmidtThomas2014JMN}
D.~R. Schmidt and P.~J. Thomas, ``Measuring edge importance: a quantitative
  analysis of the stochastic shielding approximation for random processes on
  graphs,'' {\em The Journal of Mathematical Neuroscience}, vol.~4, no.~1,
  p.~6, 2014.

\bibitem{kolmogorov1956shannon}
A.~Kolmogorov, ``On the {Shannon} theory of information transmission in the
  case of continuous signals,'' {\em IRE Transactions on Information Theory},
  vol.~2, no.~4, pp.~102--108, 1956.

\bibitem{pinsker1964information}
M.~Pinsker, ``Information and information stability of random variables and
  processes. {Translated} and edited by {Amiel Feinstein}, {Holden-Day, Inc.,
  San Francisco, Calif},'' {\em London-Amsterdam xii}, vol.~243, 1964.
\newblock Cf.~Equation 10.4.1.

\bibitem{ihara1993information}
S.~Ihara, {\em Information theory for continuous systems}, vol.~2.
\newblock World Scientific, 1993.

\bibitem{nips:ThomasEtAl:2004}
P.~J. Thomas, D.~J. Spencer, S.~K. Hampton, P.~Park, and J.~Zurkus, ``The
  diffusion-limited biochemical signal-relay channel,'' in {\em Advances in
  NIPS}, 2004.

\bibitem{SchmandtGalan2012PRL}
N.~T. Schmandt and R.~F. Gal{\'a}n, ``Stochastic-shielding approximation of
  markov chains and its application to efficiently simulate random ion-channel
  gating,'' {\em Physical review letters}, vol.~109, no.~11, p.~118101, 2012.

\bibitem{BernardiLindner2014JNPhys}
D.~Bernardi and B.~Lindner, ``A frequency-resolved mutual information rate and
  its application to neural systems,'' {\em Journal of neurophysiology},
  vol.~113, no.~5, pp.~1342--1357, 2014.

\bibitem{tai2003finite}
K.~Tai, S.~D. Bond, H.~R. MacMillan, N.~A. Baker, M.~J. Holst, and J.~A.
  McCammon, ``Finite element simulations of acetylcholine diffusion in
  neuromuscular junctions,'' {\em Biophysical journal}, vol.~84, no.~4,
  pp.~2234--2241, 2003.

\bibitem{bigland1983changes}
B.~Bigland-Ritchie, R.~Johansson, O.~C. Lippold, S.~Smith, and J.~J. Woods,
  ``Changes in motoneurone firing rates during sustained maximal voluntary
  contractions.,'' {\em The Journal of physiology}, vol.~340, no.~1,
  pp.~335--346, 1983.

\bibitem{eckford2018thermodynamic}
A.~W. Eckford, B.~Kuznets-Speck, M.~Hinczewski, and P.~J. Thomas,
  ``Thermodynamic properties of molecular communication,'' in {\em 2018 IEEE
  International Symposium on Information Theory (ISIT)}, pp.~2545--2549, IEEE,
  2018.

\end{thebibliography}


\end{document}